# Spatially incoherent illumination interferometry: a PSF almost insensitive to aberrations

Peng Xiao, Mathias Fink, and A. Claude Boccara*
*Institut Langevin, ESPCI ParisTech, PSL Research University, 1 rue Jussieu, 75005 Paris, France*
*Corresponding author: claude.boccara@espci.fr*

**We show that with spatially incoherent illumination, the point spread function width of an imaging interferometer like that used in full-field optical coherence tomography (FFOCT) is almost insensitive to aberrations that mostly induce a reduction of the signal level without broadening. This is demonstrated by comparison with traditional scanning OCT and wide-field OCT with spatially coherent illuminations. Theoretical analysis, numerical calculation as well as experimental results are provided to show this specific merit of incoherent illumination in full-field OCT. To the best of our knowledge, this is the first time that such result has been demonstrated.**

Aberrations can degrade the performances of optical imaging systems. This issue is particularly crucial when imaging biological samples since scattering media or multi-scale aberrating structures usually hinder the objects of interest. Aberrations are known to blur optical images by perturbing the wavefronts; more precisely the distorted optical images are obtained by amplitude or intensity convolution of the diffraction limited images with the aberrated point spread function (PSF). Depending on the nature of the illumination, spatially coherent or incoherent, intensity or amplitude has to be considered [1,2]. In order to reduce or to avoid blurring, adaptive optics (AO), which was originally proposed and developed for astronomical imaging [3,4], is usually used to correct the perturbed wavefront thus achieving diffraction-limited PSF during imaging.

Optical interferometry techniques have been widely used for imaging. Among those techniques, the use of optical coherence tomography (OCT) has increased dramatically in various researches and clinical studies since its development. Traditional scanning OCT selects ballistic (more precisely singly backscattered) photons through scattering media based on a broadband light source and coherent cross-correlation detection [5]. Both longitudinal [6,7] and *en face* scanning [8,9] OCTs use spatially coherent illumination and rely on point-by-point scanning to acquire three-dimensional reflectivity (back-scattering) images. Parallel OCT systems that take images with planes that are perpendicular to the optical axis have also been developed with specific detectors and methods by using either spatially coherent illumination like wide-field OCT [10-12] or spatially incoherent illumination like full-field OCT [13]. Higher resolutions are achieved in these systems as en face acquisition allows using larger numerical aperture optics. Wide-field OCT systems with powerful laser sources or superluminescent diodes give high sensitivity but the image can be significantly degraded by coherent cross-talks [14]. Full-field OCTs use thermal lamps or light-emitting diodes for high resolution, highly parallel image acquisitions but could suffers low power per spatial mode [15].

In this paper, we show that with spatially incoherent illumination, the resolution of full-field OCT is almost insensitive to aberrations. Instead of considering the PSF of a classical imaging system such as a microscope, we will pay attention to the system PSF of interferometric imaging systems for which an *undistorted* wavefront from a reference beam interferes with the *distorted* wavefront of the object beam. More precisely we will consider the cases of scanning OCT with spatially coherent illumination, wide-field OCT with spatially coherent illumination and full-field OCT with spatially incoherent illumination; surprisingly we found that in full-field OCT with incoherent illumination the system PSF width is almost independent of the aberrations and that only its amplitude varies.

In order to stick to the PSF definition, we will consider a point scatterer as our object and will analyze the response of the system to such object. Suppose the single point scatterer is at position $(x', y') = (a, b)$, the sample arm PSF of the interferometer is $h_s$ and the reference arm PSF of the interferometer is $h_r$. For simplification, we ignore all the constant factors in the following expressions. So in all the three cases, the sample field at the detection plane would be

$$g_s = h_s(x' - a, y' - b) \quad (1)$$

In the case of traditional scanning OCT, the reference field of each scanning position at the detection plane would be $h_r(x - x', y - y')$. Since coherent illumination is used, Interference happens at each scanning position and the final interference would be a sum of the interference term across the scanning filed result in

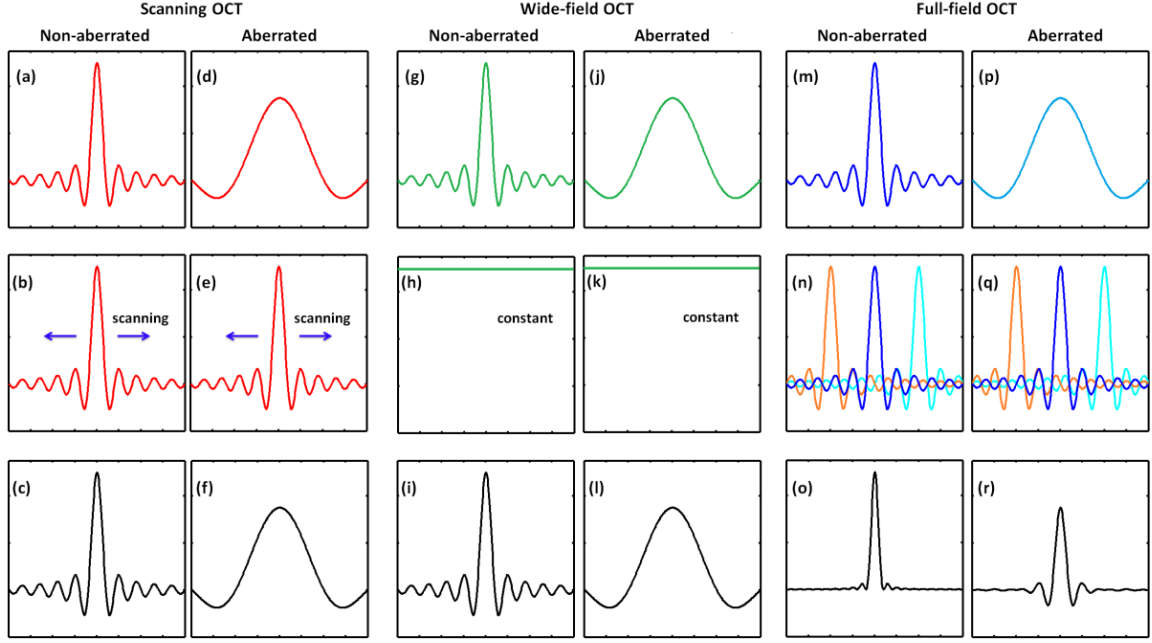

Fig. 1. Illustration of single point scatterer (PSF) interferences in both non-aberrated and aberrated sample arm PSF situations for scanning OCT and wide-field OCT with spatially coherent illumination and FFOCT with spatially incoherent illumination. (a,g,m) Non-aberrated sample arm PSF, (d,j,p) Aberrated sample arm PSF, (b,e) Scanning reference arm PSF for scanning OCT, (h,k) Constant reference field for wide-field OCT, (n,q) Reference arm PSFs for full-field OCT, (c,f,i,l,o,r) The corresponding interference signal (system PSF). Different colors in (n,q) indicate the spatial incoherence from each other.

$\langle g_s g_r \rangle_s = \iint h_s(x'-a, y'-b) h_r(x-x', y-y') dx' dy'$ (2)

Thus, the system PSF of scanning OCT system would be a convolution of the sample arm PSF and the reference arm PSF as shown in Fig. 1 (a-c). When aberrations exist, the convolution of the aberrated sample arm PSF with the diffraction-limited reference arm PSF results in an aberrated system PSF for the scanning OCT systems (Fig. 1 (d-f)).

In the case of wide-field OCT, as coherent sources are used, the optical beams are typically broadened by lenses to form parallel illuminations on both arms of the interferometer [12]. Thus plane waves impinge on both the object and the reference mirror. In the sample arm, the point scatterer will send a spherical wave back that will be focus on the camera plane that can be described by expression (1). For the reference arm, consider it as homogeneous illumination, a plane wave will be reflected back by the reference mirror and form a uniform field at the camera plane. Thus the interference happen between the two arms would be

$\langle g_s g_r \rangle_w = h_s(x'-a, y'-b)$ (3)

as constant value is ignored. So the system PSF is actually defined by the sample PSF. It is illustrated in Fig. 1 (g-i). When aberrations distort the backscattered wavefront of the sample arm, the aberrated sample arm PSF interferes with a uniform reference field results in an aberrated system PSF for the wide-filed OCT systems (Fig. 1 (j-l)).

When we deal with the case of full-field OCT with spatially incoherent illumination, we have to go back to the basic definition of the spatial coherence of the beams that impinge the reference arm as well as the sample arm of the interferometer. Let's consider a circular uniform incoherent source located in the image focal plane of a microscope objective with a focal length of $f_0$, which could be obtained with a standard Koehler illumination. The source illuminates the field of view of the microscope objective.

One first step is to determine the spatial coherence length in the field of view. The Van Cittert-Zernike theorem states that the coherence angle is given by the Fourier transform of the source luminance [16]. If the pupil diameter is $D$, the angle would be defined as $\sin \alpha = \lambda/D$. At the level of focal plane, this corresponds to a zone of radius $\rho = f_0 \lambda/D$ or $\rho = \lambda/2NA$. We can say that, in absence of aberrations, the focal plane is "paved" by small coherent areas (CA) of radius $\rho$. This radius is also the radius of the diffraction spot that limits the resolution of the microscope objective in absence of aberrations. When going from one diffraction spot to the next adjacent diffraction spots the incoherent plane waves impinging the objective are separated by $\pm \lambda$ on the edges of the pupil.

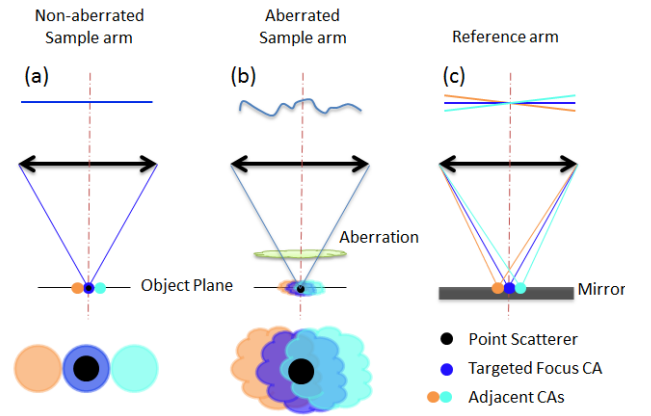

Fig. 2. Illustration of the sample and reference wavefronts in spatially incoherent interferometer with a single point scatterer in cases of non-aberrated and aberrated sample arm. Different colors in CAs and wavefronts indicate different spatial modes.

In absence of aberrations for an interferometry like full-field OCT, the single point scatterer at the object plane of the sample arm lies in a single CA (Fig. 2(a)) and the backscattered signal will only interfere with signal reflected from the corresponding CA in the reference arm (Fig. 2(c)). Note that the size of the CAs is the same as the diffraction spot, the signal from one CA at the camera plane could be expressed as the reference PSF. Thus the interference would be

$$\langle g_s g_r \rangle_f = h_s(x' - a, y' - b) h_r(x' - a, y' - b) \quad (4)$$

The system PSF is actually the dot product of the sample PSF and the reference PSF as shown in Fig. 1 (m-o). The overall signal reflected from the reference mirror at the camera is still homogenous but we displayed it by combining multiple reference PSFs reflected from different CAs that have different spatial modes.

When aberrations exist in the sample arm, the various CAs in the object plane will have larger sizes and will overlap each other (Fig. 2(b)). This result in the backscattered signal of the single point scatterer in the sample arm containing not only the spatial mode of the targeted focus CA but also the modes from the overlapped adjacent CAs. Thus with aberrations that create a broadened sample PSF, interference will happen not only with the reference beam corresponding to the targeted CA, but also with the beams corresponding to the adjacent CAs. What we want to demonstrate and to illustrate by an experiment is that the interference signal with the targeted focus CA gives a much stronger signal than the one with the adjacent CAs resulting in an "interference" PSF that is much thinner than the one of the classical broadened sample PSF. At the level of the image plane, the interference between the sample aberrated beam and the non aberrated reference beam is only constructive in a zone limited by the spatial coherence of the reference beam. In order to be more quantitative we are going to compare this by the Strehl ratio approach.

The "best focus" signal intensity damping compared to the diffraction limited PSF is given (for small aberrations) by the Strehl ratio that is proportionnal to the peak aberrated image intensity $S = e^{-\sigma^2}$, where $\sigma$ is the root mean square deviation over the aperture of the wavefront phase $\sigma^2 = (std(\phi))^2$. In our case, suppose $\phi$ is the phase of the interference wavefront between the sample signal and the reference signal corresponding to the targeted focus CA, then the phase of the interference wavefront with the reference signals corresponding to an adjacent CAs is $\phi + \phi_1$, where $\phi_1$ is a phase that varies linearly from one edge of the pupil to the other in the range of $\pm 2\pi$. A comparison between the signal ratio of the interference signal with the targeted CA and the one with an adjacent CAs

$$s_t = e^{-(std(\phi))^2} \gg s_a = e^{-(std(\phi+\phi_1))^2} \quad (5)$$

shows that the influence of off axis Cas is negigeable $s_a$. Let's consider various aberrations leading to a significant Strehl ratio of 0.03, numerical calculations results are shown in Fig. 3. For defocus, the intensity ratio of the interference with adjacent CAs is damped for about 740 times compared with the interference with the targeted focus CA, resulting in a signal damping or an amplitude damping of 27.1 times. The amplitude damping ratio is calculated by

$$Amplitude\ damping\ ratio = \sqrt{s_t/s_a} \quad (6)$$

as amplitude instead of intensity is obtained in full-field OCT signal. It's easy to prove that this value is fixed for all the axisymmetric aberrations like defocus, astigmatism, spherical aberrations, etc. While for coma with a Strehl ratio of 0.03, the simulated amplitude damping ratio is $8.2 - 86.1$ times depending on the spatial position of the adjacent CAs. In another word, the interference signal was severally damped going from the targeted CA to the adjacent CAs. Thus in the camera plane, as shown in Fig 1. (p-r), the interference signal result in a dot product of the aberrated sample PSF with the reference PSF corresponding to the targeted focus CA since the interference with the reference PSFs corresponding to the adjacent CAs are significantly degraded. This actually matches with equation (4) for non-aberrated situation, the system PSF could be calculated by the dot product of the sample PSF and the reference PSF. For distorted sample PSF (mostly broadened), its interference with the reference channel conserves the main feature of an unperturbed PSF with only a reduction in the FFOCT signal level. We mentioned "almost" for the resolution conservation, because there are situations in which the product of the reference arm PSF with off-center aberrated sample arm PSF may results in losing some sharpness due to the high side lobes of the Bessel PSF function.

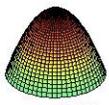

| Zernike Mode | Targeted Focus wavefront | Adjacent Wavefront | Amplitude Damping Ratio |
|---|---|---|---|
| Defocus | $s_t = 0.03$ | $s_a = 4.07 \times 10^{-5}$ | 27.1 |
| Astigmatism | $s_t = 0.03$ | $s_a = 4.07 \times 10^{-5}$ | 27.1 |
| Coma | $s_t = 0.03$ | $s_a = 4.07 \times 10^{-5}$ | 8.2 - 86.1 |
| Spherical Aberration | $s_t = 0.03$ | $s_a = 4.07 \times 10^{-5}$ | 27.1 |

Fig. 3. Aberrated interference wavefronts and numerical simulations of the Strehl ratio and amplitude damping for interference with targeted CA and adjacent CAs. Defocus, astigmatism, coma and spherical aberration are considered. The damping for coma varies depending on the spatial position of the adjacent CAs.

With the commercial LLtech full-field OCT system Light-CT scanner [17], we have also conducted experiments with gold nanoparticles to check how the system PSF would be affected by inducing different level of defocus. 40nm radius gold nanoparticles solution was diluted and dried on a coverslip so that single particles could be imaged. By moving the sample stage, 10um, 20um and 30um defocus was induced to the targeted particle. The length of the reference arm was shifted for the same value in order to match the coherence plan of the two arms for imaging. Theoretically, the system resolution was 1.5um corresponding to about 2.5 pixels on the camera. By adding 10um, 20um and 30um defocus, the sample PSF would be broadened by 2.3 times, 4.6 times and 6.9 times. Experimental results are shown in Fig. (4).

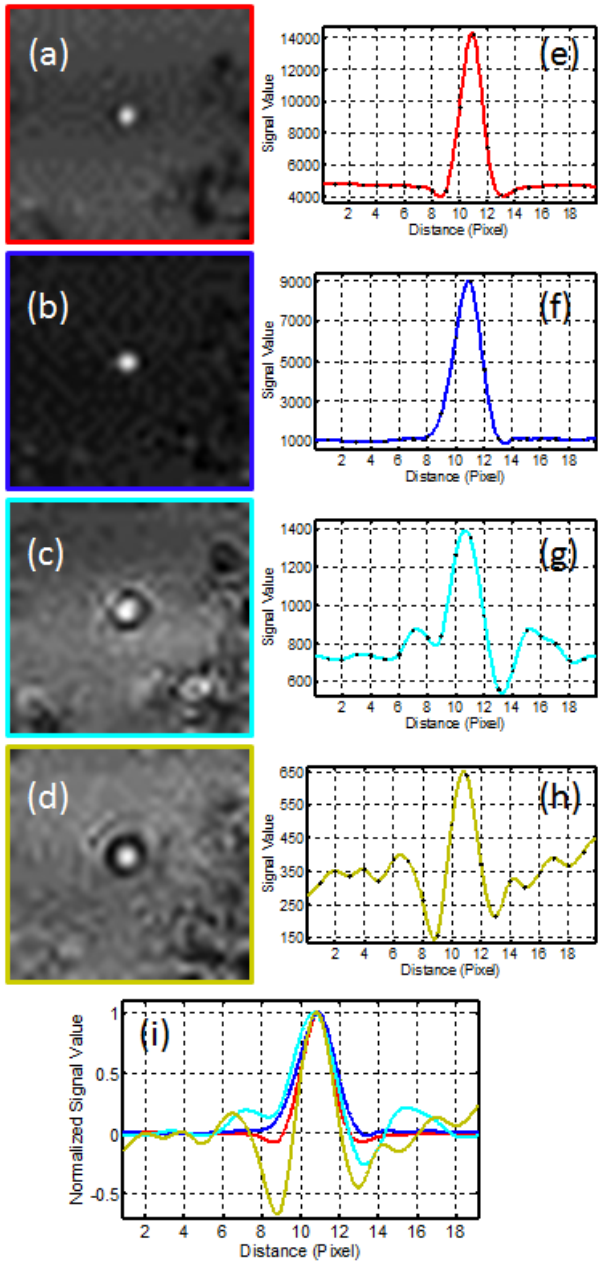

Fig. 4. Full-field OCT experiment results of gold nanoparticle by adding different level defocus. Full-field OCT images (a-d) and the corresponding intensity profile (e-h) of a targeted nanoparticle are shown for well-focused (a,e) and defoused for 10 um (b,f), 20 um (c,g) and 30 um (d,h) situations. Normalized PSF profiles are shown in (i) indicating no obvious broadening are observed after inducing different level of defocus.

Full-field OCT images (Fig. 4(a-d)) and the corresponding signal profiles (Fig. 4(e-h)) of the same nanoparticle were displayed. It's obvious that with more defocus added the signal level of the gold nanoparticles is reduced, but the normalized signal profiles graph (Fig. 4(i)) shows clearly that the size of the particle that corresponds to the system PSF width keeps the same for all the situations.

In conclusion, we have shown for the first time to our best knowledge that in spatially incoherent illumination interferometry like full-field OCT, the system PSF width is almost insensitive to aberrations with only signal amplitude reduction. This is demonstrated by a simple theoretical analysis as well as numerical simulations for different aberrations, and confirmed by experiments with a full-field OCT system. More precisely the aberration-induced reduction in signal is roughly proportional to the square root of the Strehl ratio. Let us consider the realistic case of a diffraction-limited imaging system with a PSF width of 2 μm that allows for instance resolving the cones in retinal imaging. With a Strehl ratio of 0.1, which is considered to give a low quality image, the PSF would be broadened to about 6 μm that would mask the cell structures. But in full-field OCT system, the same Strehl ratio would only reduce the signal by a factor of 3.1 while keeping the image sharpness.

As we intended to apply full-filed OCT system with adaptive optics for eye examination, this specific merit of spatially incoherent illumination could simplified the in vivo observation of the eye. We think that we could restrict the aberration corrections to the main aberrations (e.g. focus and astigmatism) that will improve the signal to noise ratio and skip the high order aberrations. This would also increase the correction speed thus reducing the imaging time. A large number of experiments of USAF resolution target and biological samples with induced or natural aberrations have confirmed that the resolution is maintained and only signal-to-noise ratio was degraded. These results will be submitted soon.